\documentstyle[aps,overcite,graphics,multicol,array,epic,eepic,amsmath]{revtex}

\begin{document}

\def\M{{\bf M}}
\def\A{{\bf A}}
\def\B{{\bf B}}
\def\F{{\bf F}}
\def\X{{\bf X}}
\def\H{{\bf H}}
\def\R{{\bf R}}
\def\S{{\bf S}}
\def\U{{\bf U}}
\def\u{{\bf u}}
\def\t{{\bf t}}
\def\x{{\bf x}}
\def\ab{{\bf a}}
\def\bb{{\bf b}}
\def\eb{{\bf e}}
\def\fb{{\bf f}}
\def\gb{{\bf g}}
\def\lb{{\bf l}}
\def\yb{{\bf y}}
\def\zb{{\bf z}}
\def\db{{\bf d}}
\def\zero{{\mbox{\bf $0$}}}
\def\kk{{\bf k}}
\def\vv{{\bf v}}
\def\qq{{\bf q}}
\def\rr{{\bf r}}
\def\pp{{\bf p}}          
\def\P{{\bf P}}
\def\E{{\bf E}}
\def\I{{\bf I}}
\def\J{{\bf J}}
\def\T{{\bf T}}
\def\Pt{{\tilde{P}}}
\def\Qt{{\tilde{Q}}}
\def\Li{{\cal{L}}}
\def\gradrn{{\nabla_{{\bf{R}}^{N}}}}
\def\gradrj{{\nabla_{{\bf{R}}_{j}}}}
\def\gradpn{{\nabla_{{\bf{P}}^{N}}}}
\def\gradpj{{\nabla_{{\bf{P}}_{j}}}}
\def\odagger{{\cal O}^{\dagger}}
\def\btau{\mathbf{\tau}}
\def\bphi{\mbox{\boldmath{$\phi$}}}
\def\btheta{ \mbox{\boldmath{$\theta$}}  } 
\def\hP{{\hat{P}}}
\def\hp{{\hat{p}}}
\def\hq{{\hat{q}}}
\def\hQ{{\hat{Q}}}
\def\hB{{\hat{B}}}
\def\hH{{\hat{H}}}
\def\Lo{{{\cal L}}}
\def\ap{{\alpha^{\prime}}}
\def\bp{{\beta^{\prime}}}
\def\a{{\alpha}}
\def\b{{\beta}}
\def\at{{\tilde{\alpha}}}
\def\atp{{\tilde{\alpha}^{\prime}}}
\def\bt{{\tilde{\beta}}}
\def\btp{{\tilde{\beta}^{\prime}}}
\def\mut{{\tilde{\mu}}}
\def\nut{{\tilde{\nu}}}
\def\hmu{{\hat{\mu}}}
\def\hnu{{\hat{\nu}}}

\def\M{{\bf M}}
\def\A{{\bf A}}
\def\B{{\bf B}}
\def\F{{\bf F}}
\def\X{{\bf X}}
\def\H{{\bf H}}
\def\R{{\bf R}}
\def\S{{\bf S}}
\def\U{{\bf U}}
\def\u{{\bf u}}
\def\t{{\bf t}}
\def\x{{\bf x}}
\def\ab{{\bf a}}
\def\bb{{\bf b}}
\def\eb{{\bf e}}
\def\fb{{\bf f}}
\def\gb{{\bf g}}
\def\lb{{\bf l}}
\def\yb{{\bf y}}
\def\zb{{\bf z}}
\def\db{{\bf d}}
\def\zero{{\mbox{\bf $0$}}}
\def\kk{{\bf k}}
\def\vv{{\bf v}}
\def\qq{{\bf q}}
\def\rr{{\bf r}}
\def\pp{{\bf p}}          
\def\P{{\bf P}}
\def\E{{\bf E}}
\def\I{{\bf I}}
\def\J{{\bf J}}
\def\T{{\bf T}}
\def\Pt{{\tilde{P}}}
\def\Qt{{\tilde{Q}}}
\def\Li{{\cal{L}}}
\def\gradrn{{\nabla_{{\bf{R}}^{N}}}}
\def\gradrj{{\nabla_{{\bf{R}}_{j}}}}
\def\gradpn{{\nabla_{{\bf{P}}^{N}}}}
\def\gradpj{{\nabla_{{\bf{P}}_{j}}}}
\def\odagger{{\cal O}^{\dagger}}
\def\btau{\mathbf{\tau}}
\def\bphi{\mbox{\boldmath{$\phi$}}}
\def\btheta{ \mbox{\boldmath{$\theta$}}  } 
\def\hP{{\hat{P}}}
\def\hp{{\hat{p}}}
\def\hq{{\hat{q}}}
\def\hQ{{\hat{Q}}}
\def\hB{{\hat{B}}}
\def\hH{{\hat{H}}}
\def\Lo{{{\cal L}}}
\def\ap{{\alpha^{\prime}}}
\def\bp{{\beta^{\prime}}}
\def\a{{\alpha}}
\def\b{{\beta}}
\def\at{{\tilde{\alpha}}}
\def\atp{{\tilde{\alpha}^{\prime}}}
\def\bt{{\tilde{\beta}}}
\def\btp{{\tilde{\beta}^{\prime}}}
\def\mut{{\tilde{\mu}}}
\def\nut{{\tilde{\nu}}}
\def\hmu{{\hat{\mu}}}
\def\hnu{{\hat{\nu}}}

\title{An efficient Monte Carlo method for calculating ab initio
transition state theory reaction rates in solution}
\author{Radu Iftimie${}^{1,3}$, Dennis Salahub${}^{2}$ and Jeremy Schofield${}^{1}$}
\address{1. Chemical Physics Theory Group,
Department of Chemistry, University of Toronto,
Toronto, Ontario, Canada M5S 3H6}
\address{2. University of Calgary,
2500 University Drive N.W.,
Calgary, Alberta, Canada T2N 1N4}
\address{3. New York University, Department of Chemistry,
100 Washington Sq. E., New York, NY, 10003
}

\date{\today}
\maketitle
\begin{abstract}
In this article, we propose an efficient method for sampling the 
relevant state space in condensed phase reactions. In the present method,
the reaction is described by solving
the electronic Schr\"{o}dinger equation for the solute atoms in the presence 
of explicit solvent molecules. The sampling algorithm uses a molecular mechanics
guiding potential in combination with simulated tempering ideas and allows thorough
exploration of the solvent state space in the context of an ab initio
calculation even when the dielectric
relaxation time of the solvent is long. The method is applied to the study of 
the double proton transfer reaction that takes place between a molecule of acetic
acid and a molecule of methanol in tetrahydrofuran. 
It is demonstrated that calculations
of rates of chemical transformations occurring in solvents of medium
polarity can be performed with an increase in the cpu time of factors ranging from
4 to 15 with respect to gas-phase calculations.

\end{abstract}

\section{Introduction}

The concept of reaction mechanism plays a major role in chemistry
and represents a synthesis of our understanding of the way in which
different topological changes in the bonding structure of a reactant
or product are correlated as the reaction proceeds. 
Recent advances in ultrafast lasers\cite{FlemingPhysToday90,ZewailExperimental},
X-ray\cite{PermanScience98} and other spectroscopies as well as in
computational chemistry\cite{RosskyNature94,ClaryTheoreticalGasPhase}
have made possible the study of most gas-phase and some condensed-phase
reactions in molecular detail. However, most experimental investigations
of complex reaction mechanisms taking place in liquid environments
are still inferred from isotope and solvent (medium) effects on the
reaction rate\cite{KlinmanBiophys.,PageWilliams}. Consequently, the
interpretation of the experimental results as well as the reaction
mechanisms inferred are more controversial than those of gas-phase
reactions. 

Computer studies can be useful as a complement to experimental data
in cases where experiments alone cannot provide a definitive picture
of the mechanism of the chemical process. It is therefore desirable to
develop systematic computational approaches to carefully examine the relation between
isotope effects and reaction mechanism in condensed phase systems.
However, computational calculations of kinetic isotope effects in
condensed-phase reactions can become expensive due to a number of
difficulties. Some of the practical challenges involving the calculation
of kinetic isotope effects and reaction rates in solution are
associated with the fact that accurate descriptions of transformations in which chemical
bonds are broken and formed require time-consuming \emph{ab initio}
electronic structure methods. Computer time limitations become particularly
relevant when investigating ``rare events'' such as
chemical reactions, especially when the reactions are accompanied
by substantial differences in the structure of the solvent. To further
complicate matters, quantum effects such as zero-point vibrations and tunneling
effects are important in some chemical processes, such as proton transfer reactions.
Another technical problem in the simulations of reactive systems is
that the statistical resolution 
of calculations of the reaction rate depends on how many statistically
independent configurations are obtained during the simulation:
Simulations in which a large number of successive configurations have
similar configurations of the reactive core or of the solvent
molecules suffer from large uncertainties in the
calculated reaction rates, precluding any definitive interpretation of
the reaction mechanism.

It is therefore critical to develop methods which sample statistically
independent configurations along the reaction path rapidly and correctly.
In the case where the reaction path can be characterized by means
of a small number of reaction coordinates,
accurate, statistically well-resolved calculations of reaction rates
can be performed by developing improved methods for computing reaction
free energy profiles along these reaction coordinates.
A number of techniques for computing free energy profiles along reaction
coordinates have been proposed in the literature, including umbrella
sampling\cite{UmbrellaSampling}, thermodynamic integration in conjunction
with the blue-moon ensemble method\cite{BlueMoonMethod}, projection
methods\cite{ProjectionMethods}, variable transformation approaches\cite{VariableTransform}
and guiding potentials\cite{Iftimie1,Rothlisberger2000}. The use
of molecular mechanics-based guiding potentials was proposed simultaneously
and independently by Iftimie et al.\cite{Iftimie1} and Vondele et
al.\cite{Rothlisberger2000} and implemented in a Monte Carlo and
a molecular dynamics framework, respectively. The basic idea of the
method consists of using a ``fast'' \emph{molecular
mechanics} potential to guide a computationally intensive \emph{ab
initio} simulation. 

The Monte Carlo version of the ``guiding'' approach
in reference {[}\citeonline{Iftimie1}{]} was called the molecular
mechanics based importance function method (MMBIF). It was demonstrated
that the utilization of a reasonably accurate molecular mechanics
potential as an importance function decreases the correlation of an
\emph{ab initio} Monte Carlo calculation by two orders of magnitude.
The method was illustrated on a gas-phase formic acid-water system
in which the activated processes involved breaking and forming hydrogen
bonds\cite{Iftimie1}, and was successfully applied to calculate the kinetic isotope
effects in a model gas-phase intramolecular proton transfer
reaction\cite{Iftimie2, Iftimie3, Iftimie4}.

One of the major challenges in \emph{ab initio} simulations of reactions in condensed
phase environments is to thoroughly sample configurations of the
system when changes in the solvent occur on long time scales.  For
instance, in molecular dynamics simulations of proton transfer reactions in which the 
collective behavior of the solvent can strongly influence the dynamics
of the reaction, the sampling efficiency can be limited by long
solvent dielectric relaxation time. In essence, an independent
configuration of the system requires that the equations of motion be
propagated for a time which is longer than the dielectric relaxation
time.  Even simple organic solvents such as tetrahydrofuran, the
relevant solvent in this study, have dielectric relaxation times on
the order of $4$ ps\cite{relaxationTHF}, so that independent solvent
configurations are only obtained after several thousand elementary
propagation steps.  More structured solvents such as water, with a
dielectric relaxation time of roughly $8.3$ ps\cite{relaxationWater},
require even longer propagation for proper sampling of solvent configurations.
The long time scale of structural rearrangements in solvents pose a
serious challenge to \emph{ab initio} calculations even when the solvent is modeled using molecular
mechanics since each propagation step in the dynamics involves a
time-consuming \emph{ab initio} calculation.  Ideally, successive
configurations in a simulation involve drastically different solvent
and solute configurations.  This is only possible using an artificial
dynamics to generate the sequence of configurations.  One way to generate
relatively uncorrelated successive configurations is to apply
importance sampling ideas.

In this paper, the molecular mechanics-based importance sampling
method is adapted to calculate reaction rates of chemical processes in
condensed phases where collective motions of the environment can 
influence the quantitative features of the chemical
process and, in some cases, play a critical role in determining the
mechanism of a reaction. The Monte Carlo procedure involves separating the task of
sampling the configurations of the condensed phase system into two
parts.  The first part involves an efficient scheme of 
updating the solvent configuration while the second focuses on the
relatively slow \emph{ab initio} calculation of the reactive core. This
approach allows extensive sampling of the molecular mechanical solvent
without a significant increase in the overall computational work over
a gas phase \emph{ab initio} simulation.  The method is applied
to study the double proton transfer reaction in an acetic
acid-methanol complex solvated by tetrahydrofuran.

\section{Methodology}
\subsection{Motivation for the Model System}
The computational study of proton-transfer
reactions can be used to understand the conditions for the validity of
a well-known conjecture proposed
in the physical organic chemistry literature, stating that the breakdown
of the rule of geometric mean for kinetic isotope effects, which is a relation\cite{KohenJACS2002} involving
ratios of kinetic isotope effects corresponding to different isotopic
substitutions at the primary and secondary atoms, is a signature
of tunneling for both primary and secondary atoms\cite{KlinmanBiophys.,LimbachBook}.
The consequences of applying the rule of geometric mean to interpret
experimentally determined isotope effects are of far-reaching importance:
The inferred relationship between the rule of geometric mean and reaction
mechanism forms the basis for the proposal that \emph{multiple} \emph{intramolecular}
proton transfer reactions are likely to proceed via a two-step mechanism,
in contrast to \emph{multiple intermolecular} proton transfers which
are believed to proceed via a synchronous pathway\cite{LimbachBook}.
The same relationship is at the heart of the recent suggestion that
tunneling effects have played an important role in the design of the
active sites of some proteins\cite{KlinmanBiophys.,KlinmanJACS2001}.

Some of the most striking consequences of the rule of geometric 
mean\cite{LimbachBook,KlinmanBiophys.}
appear when studying multiple proton transfer reactions in condensed
phases. The study of reactions involving
the exchange of a pair of protons between two molecules may provide
insight into the dynamics of certain types of enzymatic reactions
in which several functional groups in the active center are properly
aligned so that concerted catalysis can occur. This type of catalysis
mechanism is called \emph{bifunctional catalysis} and is the principal
mechanism responsible for the several orders of magnitude increase
in the reaction rate in several important biochemical transformations
\cite{Dugas}. 

The double proton transfer reaction between acetic
acid and methanol is one of the simplest examples of reactions involving
an intermolecular exchange of protons between two molecules and therefore
is a good candidate for computationally investigating general aspects
of bifunctional catalysis.  The chemical processes occurring in a
solution of acetic acid-methanol in tetrahydrofuran (THF) have been
studied experimentally by Gerritzen and Limbach\cite{Limbach84}. The majority species in the
system consist of complexes formed from either a single molecule of acetic acid
or a single molecule of methanol hydrogen-bonded with a single molecule of solvent.
The minority species in the system consist of linear and
cyclic clusters of acetic acid hydrogen-bonded with methanol
and solvated by tetrahydrofuran. The double-proton transfer reaction
takes place along the hydrogen bonds of the cyclic clusters. When
the concentrations of acetic acid and methanol are reduced, the only
cyclic cluster formed within which double-proton transfer is experimentally
observed is formed from one molecule of acetic acid and a single molecule of
methanol\cite{Limbach84} (see Fig. 1).

In this work, we will focus only on
the intermolecular double proton transfer
which takes place in the cyclic cluster formed from one molecule of
methanol and a single molecule of acetic acid in a solution of
tetrahydrofuran (THF). The quantum nuclear effects due to the small mass
of the hydrogen atoms will be neglected here, and we will concentrate only
on the sampling issues. Once the sampling method is tested for classical atoms,
the method will be developed in a later article to include nuclear quantum
effects via centroid transition state theory using sampling ideas similar to those
used in our previous studies of malonaldehyde\cite{Iftimie2, Iftimie3}.

\subsection{The QM/MM/Implicit solvent approach}
Since it is impractical to treat large systems quantum mechanically,
one is inevitably faced with the decision of how to combine \emph{ab initio}
electronic structure methods with more empirical approaches.  One
alternative to full \emph{ab initio} calculations consists of using a mixed
quantum-mechanical/molecular-mechanical (QM/MM) description of the
system.  However, in
general it is problematic to clearly define the physics of the
interface region between the quantum and molecular mechanical subsystems. 
In the best of circumstances, the separation can be made without
``cutting'' a covalent bond.  For reactive processes in which the
reaction occurs in a small solute which can be simulated in isolation using \emph{ab initio} methods, 
the logical separation between the quantum and molecular mechanical
regions is at the solute-solvent level.  That is, the energies of the solute in which
the reaction is occurring are calculated using electronic structure
methods for the solute in the presence of the solvent, while the solvent energies are
described by molecular mechanical potentials.  

Even after this separation has been defined, one is faced with another
technical issue of how an infinite condensed phase system can be represented in a
practical fashion.  One common approach utilized in simulations of
condensed phase systems is to simulate explicitly a system consisting
of a solute surrounded by a small number of solvent molecules
periodically replicated in an infinite fashion.  Unfortunately, such a
periodic replication of the system introduces artifacts.  One such artifact
is the existence of spurious long-ranged correlations due to the
periodicity of the system.  In principle, such correlations could have
a large impact on the reactive process.  The use of periodic boundary
conditions also introduces complexity in the electronic structure
calculation itself when localized basis sets are used.  An alternative
approach is to use toroidal boundary conditions in which part of the
solvent is explicitly represented while the influence of the solvent
from regions far from the solute is implicitly incorporated.

In the present study we have utilized a quantum mechanics/molecular
mechanics/implicit solvent (QM/MM/continuum) approach, in which the
bond breaking and forming processes taking place in the solute are described
by including all solute electron-electron, electron-nuclei and nuclei-nuclei
interactions in an electronic Hamiltonian. The interactions between
the solvent molecules which form the first few solvation shells (in
practice, $343$ THF molecules)  are
described using a molecular mechanics potential. The electrostatic
interactions between the solute and the solvent molecules in the first
few solvation shells are incorporated into this approach by solving
the \emph{ab initio} electronic structure equations for the solute
in the presence of the electric field generated by the molecular mechanics
solvent charges. The total energy of the system also includes the
Lennard-Jones interaction energy between solute and the solvent molecules
in the first solvation shells, and the reaction field energy which
accounts for the long-range electrostatic interactions between solute
and solvent molecules (see Fig. 2). 

The simulations were performed using the
toroidal boundary conditions approach\cite{Neumann}, in which
a cutoff distance is used that sets the boundaries of the region within which
intermolecular interactions are explicitly counted.
Therefore, the only quantum electron-electron,
electron-nuclei and nuclei-nuclei interactions which were 
explicitly counted in the present treatment were
those which correspond to atoms separated by less than the cutoff
distance, here chosen to be $14$ angstroms.

\newcommand{\D}{\displaystyle} \normalsize

The effects of the neglected interactions in the toroidal boundary
conditions approach have been approximately accounted for by
adding reaction field corrections\cite{Neumann} to the electronic
energy\cite{Watts}:

\begin{equation}
\label{ReactionFieldEnergy}
E_{RF}\approx -\sum _{i}\frac{\epsilon _{r}-1}{(2\epsilon _{r}+1)R_{c}^{3}}\mu _{i}M_{i},
\end{equation}
where the sum is over all molecules i inside the primitive cell ,
\( \epsilon _{r} \) is the dielectric constant of the solvent, \( R_{c} \)
is the radius of the spherical surface, \( \mu _{i} \) is the dipole
moment of molecule \( i \), and \( M_{i} \) is the total dipole
moment inside the spherical surface surrounding the molecule \( i \). 

The correct energetics in hydrogen-bonded systems and in systems
undergoing proton transfer reactions
is difficult to describe even with \emph{ab initio} methods. In particular,
DFT studies of weak hydrogen-bonding systems have proved to be particularly
difficult and only limited success has been achieved in predicting the geometries and energies
for reactant and transition state configurations on
the potential energy surface using most exchange-correlation
functionals \cite{Barone}. Care should therefore be exercised when choosing a particular
\emph{ab initio} method to calculate proton transfer reaction rates.
The non-local exchange-correlation schemes developed by Proynov, Vela and Salahub
\cite{Proynov} have shown particular promise for the description of hydrogen-bonded
systems. Sirois et al. \cite{Sirois} have demonstrated that their kinetic-energy dependent
exchange functionals (BLAP and PLAP) performed better than all GGA options (BP86,
PP86, PW91), BLYP, or other hybrid methods (B3LYP, B3PW91) on systems involving
intra-molecular hydrogen bonds. The predictions
for equilibrium and transition state geometries as well as the energetics was
in agreement with high-quality post-Hartree-Fock calculations {[}CCSD(T) and
G2{]}\cite{Sirois}.  Specifically, for the gas-phase cyclic cluster of
acetic acid and methanol in Fig. 1, the activation energy using the
PLAP exchange correlation functional was found to be approximately
$16.4$ kcal/mol, in excellent agreement with the QCISD value of
$16.14$ kcal/mol\cite{Smedarchina}.

For the simulations described in the present work, the energies of different configurations
were carried out using a modified version of the LCGTO-DFT program deMon-KS3.4
\cite{Proynov2,deMon} using the PLAP exchange-correlation functional.
The DFT electronic structure calculations were carried
out as in reference [\citeonline{Sirois}], where the application of
DFT electronic structure methods to hydrogen-bonding systems is
discussed in detail.  A double-\( \zeta  \) plus polarization (DZVP) orbital basis set was
used for all atoms and the convergence level for the SCF (self-consistent field)
energy using the auxiliary fitting basis sets \cite{Sirois} was $0.01$ kcal/mol.

\subsection{The molecular mechanics potential describing the interaction between
the solvent molecules}

In order to implement the QM/MM/continuum approach to compute reaction
rates for the double proton transfer reaction in the acetic acid-methanol
cluster solvated by tetrahydrofuran, a sufficiently accurate molecular
mechanics description of the interaction between the solvent molecules
is needed\cite{Iftimie1,Iftimie2}.  In this work, the OPLS all atom
(OPLS-AA) force field of Jorgensen et al.\cite{Jorgensen} with a modified
electrostatic interaction term has been
used to describe the interactions between THF solvent molecules.  The
modifications to the electrostatics were designed to 
improve the gas-phase distribution of the partial charges in a THF
molecule as well as to improve the description of polarization effects
in condensed phases.  Since the local electrostatic environment as
well as long ranged polarization effects can influence the
proton transfer process, it is important to properly account for the
permanent and induced charges in the solvent.  To describe all such
electrostatic phenomena, all solvent molecules have been assigned
permanent and induced charges.
It can be demonstrated\cite{Thesis} using second-order perturbation
theory that the electrostatic interaction energy between
two polarizable molecules \( A \) and \( B \), each of which carries
a set of atomic permanent and induced charges, \( Q^{I}_{\, \textrm{p}} \)
and \( Q^{I}_{\, \textrm{in}} \), with \( I=1,\cdots N \)
corresponding to the charges on molecule $A$, and \( q^{i}_{\, \textrm{p}} \)
and \( q^{i}_{\, \textrm{in}} \) , where \( i=1,\cdots n \) are the
site charges on molecule $B$, respectively,
can be written to a good approximation as :

\begin{equation}
\label{TotalInteractionEnergyPermanentPlusInducedText}
V(Q^{1}_{\, \textrm{p}},\cdots Q^{N}_{\, \textrm{p}},Q^{1}_{\, \textrm{in}},\cdots 
Q^{N}_{\, \textrm{in}},q^{1}_{\, \textrm{p}},\cdots q^{N}_{\, \textrm{p}},q^{1}_{\, \textrm{in}},
\cdots q^{N}_{\, \textrm{in}})
\approx 
\sum ^{N}_{I=1}\sum ^{n}_{i=1}
\frac{Q^{I}_{\, \textrm{p}}q^{i}_{\, \textrm{p}}}{4\pi \epsilon _{0}d_{iI}}+\frac{1}{2}
\sum ^{N}_{I=1}\sum ^{n}_{i=1}\left( \frac{Q^{I}_{\, \textrm{p}}q^{i}_{\, \textrm{in}}}{4\pi 
\epsilon _{0}d_{iI}}+\frac{Q^{I}_{\, \textrm{in}}q^{i}_{\,
\textrm{p}}}{4\pi \epsilon _{0}d_{iI}}\right) ,
\end{equation}
where $d_{iI}$ is the distance between sites $I$ and $i$ on molecules
$A$ and $B$.
In principle, the induced charge appearing on a given solvent molecule
is dependent on its local environment.  One simple way of
incorporating solvent polarization effects is to assign each solvent
molecule the same average induced charge in a ``mean-field''
fashion.  More sophisticated methods of including polarization effects
either assign site polarizabilities or use fluctuating charges
distributed at specific locations on each solvent molecule\cite{RickBerne}.  Such
methods when combined with \emph{ab initio} electronic structure methods 
either necessitate an iterative solution of the electronic structure
and fluctuating charge distributions or involve dynamical methods in
an extended Lagrangian system\cite{Roitberg}.  Unfortunately, each of these
approaches has shortcomings which make them impractical to implement
in conjunction with importance sampling Monte-Carlo methods.  

In principle, the iterative minimization of the Kohn-Sham and
fluctuating charge functionals can be implemented within a Monte-Carlo
sampling approach at the cost of additional computational effort.
However, provided the solvent is not very polarizable, the variation
of the induced charges on the solvent molecules from their mean due to
the presence of the solute is expected to be small.  For this reason,
we have utilized fixed induced charges for all solvent molecules.
This corresponds to computing ground state energies for the reactive
core $E(\rho_s ({\bf{x}}))$ based on a Kohn-Sham functional\cite{HohenbergKohn,KohnSham,Levy} 
$F[\rho_s ({\bf x})]$ which depends on the ground state electron distribution  $\rho_s({\bf x})$ 
of the solute in the presence of the fixed permanent and 
induced external solvent charges. Note that this functional includes 
the electrostatic energy of interaction between the quantum solute and
the charges on the molecular-mechanical solvent molecules.

In general, one complication must be considered when discussing
electrostatic interactions in mixed QM/MM systems that arises from the
fact that the quantum mechanical electron density can become
over-polarized by the molecular mechanical point charges due to the
absence of considerations of the Fermi repulsion between quantum and
molecular mechanical charges\cite{Rothlisberger}.  Such effects
are particularly severe when using delocalized basis sets to
represent the quantum subsystem, but are less significant when
Gaussian or other local basis sets are utilized.  In the present work,
the full Coulomb interaction potential has been used to describe
electrostatic interaction terms between the solute and the solvent
charges without any screening
modifications at short distances since all DFT calculations for the
quantum subsystem use localized basis sets.

In our study of the acetic acid-methanol system in a THF solution, 
the permanent charges have been assigned the value
\( Q^{I}_{\, \textrm{p}}=q^{I}_{\, \textrm{p}}=0.887q_{\,
\textrm{opls}} \), whereas the induced charges are set to a mean-field
value of \( Q^{I}_{\, \textrm{in}}=q^{i}_{\, \textrm{in}}=0.239q_{\textrm{opls}} \) 
for all indices \( i \) and \( I \), where $q_{\textrm{opls}}$ are the
standard charges in the OPLS force field. It can be verified that the electrostatic
interaction energy calculated using
Eq.~(\ref{TotalInteractionEnergyPermanentPlusInducedText}) 
with this set of permanent and average induced charges is precisely
the electrostatic energy calculated using the standard set of fixed
OPLS charges.  On the other hand, by assigning a mean induced charge
to each solvent molecule, the computed values for the gas-phase dipole
moment and condensed phase dielectric constant are in better agreement
with the corresponding experimental values (see Table 1).

It should be emphasized that a less satisfactory means of describing
electrostatic interactions between the solute and the solvent would
consist of calculating the electrostatic interactions between the gas
phase electron distribution with the solvent charges.  In such a
scenario, the ground state electron distribution for the solute
interacts with the solvent via a Coulomb interaction of the form
\begin{eqnarray}
V= -\frac{1}{4\pi \epsilon _{0}} \sum
^{n}_{i=1}
\int \frac{\rho _{0}({\mathbf{x}}) q^{i}}{l_i(x)}  d\mathbf{x},
\label{bareInteraction}
\end{eqnarray} 
where $\rho _{0}(\mathbf{x})$ is the gas-phase electron distribution
and $l_i(x)$ is defined to be the distance between the ith charge
$q^i$ in
the solvent and the point $x$.  This approximation corresponds to 
calculating the solute energy in the condensed phase by a zeroth order
approximation for the electronic distribution of the solute, that is,
\begin{equation}
\label{InteractionSoluteSolventCorrect}
F\left[ \rho _{s}(\mathbf{x})\right] \approx  F\left[ \rho
_{0}(\mathbf{x})\right] + V.
\end{equation}
Such a description neglects the fact
that the ground state solute electron distribution is influenced by
the presence of the solvent charges.  The influence of the solvent
charges on the ground state energy can be accounted for by
incorporating a polarization energy of the solute by the solvent.

Although such a crude level of description of the electrostatic
interactions may be incomplete, it is useful in developing importance
sampling Monte-Carlo schemes based on guiding potentials, such as the
molecular mechanics based importance function method (MMBIF) described
in the next section.
\subsection{\label{SamplingMethod}The sampling methods }

The MMBIF method\cite{Iftimie1} consists of utilizing an auxiliary Markov
chain with a known asymptotic molecular mechanical distribution to propose trial
configurations for an \emph{ab initio} based Monte Carlo simulation.
In the method, each trial configuration is obtained as the last state in a classical Markov
chain generated from the current configuration in the \emph{ab initio} simulation
using various updating schemes. The proposed configurations are then accepted or rejected in the \emph{ab initio}
chain according to the usual Metropolis-Hastings algorithm\cite{Iftimie1}.
If the previous and new trial configurations in the \emph{ab initio} MC chain are
denoted by $\x _{old}$ and $\x _{new}$ respectively, the proposed state
is accepted with the probability
$\min \{1,\exp (-\triangle \triangle E/k_{B}T)\}$, where
$\Delta \Delta E$ is defined to be
\begin{equation}
\label{acceptance}
\Delta \Delta E=(E^{DFT}(\x _{new})-
E^{cl}(\x _{new}))-(E^{DFT}(\x _{old})-E^{cl}(\x _{old})),
\end{equation}
where $E^{DFT}(\x )$ and $E^{cl}(\x )$ are the potential energies
of configuration $\x$ calculated by \emph{ab initio}
methods (here density functional theory, abbreviated DFT)
and the classical potential, respectively, and $k_{B}$ is Boltzmann's constant. 
It is straightforward to show that
this acceptance criterion guarantees that the \emph{ab initio} Markov chain
has the correct limiting Boltzmann distribution\cite{Iftimie1}, regardless of the number of
classical updates used to generate the proposed configuration.

The efficiency of the MMBIF approach relies on constructing a
molecular mechanics potential for the entire system which approximates
the true interactions of the system at a qualitative level.  At first
glance, the construction of a molecular mechanics potential for a
condensed phase system appears a daunting task given that the electron
distribution of the solute changes considerably during the reactive
process and is influenced in a complicated fashion by its local
environment.  However, it is relatively straightforward to construct a
molecular mechanics potential based on simple approximate forms of the
interaction potentials, such as that in
Eq.~(\ref{InteractionSoluteSolventCorrect}).  For such forms of the
potential, the construction of the potential is reduced to modeling
the reaction in gas phase and calculating effective charges on the
gas-phase solute which mimic the correct electron distribution.  The
approximate form for the potential can be corrected for using
importance sampling methods.  For example, a Monte-Carlo chain of
states generated using an approximate expression for the energy can be
manipulated by re-weighting the configurations appearing in the chain
by an appropriate factor\cite{Iftimie2}.  The efficiency of this approach is highly
dependent on the quality of the approximation for the true energy of
the system.  One might anticipate that the crude of level of
description of the electrostatic interactions in
Eq.~(\ref{InteractionSoluteSolventCorrect}) which neglects any
polarization effects of the solute by the solvent molecules would
introduce large statistical uncertainties at the re-weighting step.
However, the polarization of the solute by the solvent can be
approximated by adjusting the charge of the solvent molecules in the
expression for the interaction between the solute and the solvent.
As will be discussed, the effective charge on the solvent molecules can
be designed to approximate solute polarization effects and thereby
improve the statistical resolution of the re-weighting procedure. 

The task of constructing a molecular mechanics potential for the
gas-phase proton transfer reaction is facilitated by using bond
evolution theory considerations.  Following these lines, 
the molecular mechanics description of the acetic acid-methanol complex 
in the absence of the solvent was created as suggested in reference {[}\citeonline{Iftimie2}{]}.
The total molecular mechanics energy was decomposed into two components.
The first component of the total energy was written as a sum of harmonic
potentials representing the variation of the potential with bond length,
bond angle or bond dihedral displacements from their minimum energy
values at a fixed value of the control parameter \( b \), defined by
\begin{equation}
\label{a_b_1andb_2Definitions}
b=d_{\textrm{O}_{7}\textrm{H}_{4}}-d_{\textrm{O}_{7}\textrm{H}_{6}},
\end{equation}
where the numbering of the atoms is that from Fig. 3.

The second component of the total energy was written as a sum of four
Morse potentials depending on the four \( \textrm{O}-\textrm{H} \)
bond lengths, plus two effective potentials depending on two parameters,
\( a_{1} \) and \( a_{2} \) defined as
\begin{equation}
\label{DefinitionAs}
\textrm{ }a_{1}=d_{\textrm{O}_{3}\textrm{O}_{7}}\textrm{ and }a_{2}=d_{\textrm{O}_{5}\textrm{O}_{7}}.
\end{equation}
These effective potentials account for the flow of electronic charge during
the reaction, as well as for the Fermi repulsion between the oxygen
atoms \( \textrm{O}_{3} \) and \( \textrm{O}_{7} \) , and between
\( \textrm{O}_{5} \) and \( \textrm{O}_{7} \). This second component
of the total energy was implemented using the same functional forms
as in reference {[}\citeonline{Iftimie2}{]}. As in the case of the
malonaldehyde study, no potential which depends explicitly on the
angles \( \widehat{\textrm{O}_{3}\textrm{H}_{4}\textrm{O}_{7}} \)
or \( \widehat{\textrm{O}_{5}\textrm{H}_{6}\textrm{O}_{7}} \) was
utilized, although some dependence on the angle \( \widehat{\textrm{H}_{4}\textrm{O}_{7}\textrm{H}_{6}} \)
was explicitly introduced using a functional form which interpolates
between a harmonic potential for transition state values of the parameter
\( b \), and zero for values of \( b \) characteristic for the reactant
or product configurations. The complete details for the construction of
the guiding potential for the solute can be found in reference {[}\citeonline{Thesis}{]}.

The simplest practical means of incorporating the electrostatic interactions
between the solute and the solvent molecules in the molecular
mechanics potential is to fit partial charges to atomic sites in the
solute to reproduce the gas-phase electronic distribution.  However,
since the electron distribution of the solute varies appreciably with
the configuration of the solute, the guiding potential must incorporate
solute charges which vary as the reaction proceeds.
In their bond evolution theory analysis of the tautomerization of
malonaldehyde, Krokidis et al.\cite{Krokidis} found that the
total charge in the basins of attraction of the proton and oxygen
atoms varies approximately linearly with a control parameter similar
to the parameter \( b \). 
Therefore, it can reasonably be assumed that most of the variation of the
charges on atomic sites in the solute can be explained by a linear
variation with \( b \).  

An alternative approach to incorporate the solute-solvent interactions
can be constructed using simulated tempering
methods\cite{MarinariParisi,Geyer1995}.  The advantage of this procedure
is that it does not rely on any approximation for the
variation of the fitted charges on the solute during the reaction.
The method essentially consists of using an extended state space to 
gradually turn on electrostatic interactions between the solute and
the solvent.  This approach is represented schematically in Figure 4.

In the simulated tempering method, a Markov chain is constructed whose
states (\( i,\mathbf{r} \)) are defined on the space formed by the
direct product between a finite set of ``electrostatic'' indices, \(
i=1,2,\cdots , i_{max} \)
and the entire solvent plus the solute configurational space. In vector notation,
the states \( \mathbf{r} \) will be denoted \( \mathbf{r}=(\mathbf{r}^{s},\mathbf{r}^{S}) \),
where \( \mathbf{r}^{s} \) and \( \mathbf{r}^{S} \) represent the
solute and solvent degrees of freedom, respectively. This Markov chain,
whose generic state is denoted (\( i,\mathbf{r} \)), is constructed
to produce states asymptotically distributed according to the
probability density 

\begin{equation}
\label{p(i,x)}
p(i,{\mathbf{r}})=w_{i}p_{i}({\mathbf{r}}),
\end{equation}
where \( w_{i} \) are constants which will be referred to as the
weights for the unnormalized probability densities \( p_{i}({\mathbf{r}}) \) defined by

\begin{equation}
\label{p_i(x)}
p_{i}({\mathbf{r}}) = \exp \left( -\beta E_{i}({\mathbf{r}})\right) ,
\end{equation}
where $\beta=1/k_{B}T$.
The potentials \( E_{i}({\mathbf{r}}) \) contain five components: the
\emph{ab initio} potential \( E^{s}({\mathbf{r}}^{s}) \) calculated
for the gas-phase solute configuration, the molecular mechanics potential
\( E^{S}(\mathbf{r}^{S}) \) describing the interaction between the
explicit solvent atoms, the Lennard-Jones potential \( E_{LJ}^{sS}({\mathbf{r}}^{s},{\mathbf{r}}^{S}) \)
describing the dispersive and short-ranged interactions between solute
and solvent atoms, the Coulomb electrostatic interaction \( E_{i}^{sS}({\mathbf{r}}^{s},{\mathbf{r}}^{S}) \)
between some charges on the solute atoms and the permanent and induced
charges on the solvent atoms, and the reaction field energies
$E_{RF}({\mathbf{r}}^{s},{\mathbf{r}}^{S})$ describing
the long-range solute-solvent electrostatic interactions:

\begin{equation}
\label{E_i(x)}
E_{i}({\mathbf{r}}^{s},{\mathbf{r}}^{S})=E^{s}({\mathbf{r}}^{s})+E^{S}({\mathbf{r}}^{S})
+E_{LJ}^{sS}({\mathbf{r}}^{s},{\mathbf{r}}^{S})+
E_{i}^{sS}({\mathbf{r}}^{s},{\mathbf{r}}^{S})+E_{RF}({\mathbf{r}}^{s},{\mathbf{r}}^{S}).
\end{equation}

The Coulomb solute-solvent electrostatic interaction potential between
charges $q^s$ on the solute and $q^S$ on the solvent, given by
\begin{equation}
E_{i}^{sS}({\mathbf{r}}^{s},{\mathbf{r}}^{S}) = 
\lambda_i \sum^{N}_{J=1}\sum ^{n}_{j=1}
\frac{q_{J}^{S} q_{j}^{s}}{4\pi \epsilon _{0}d_{Jj}},
\label{electroSoluteSolvent}
\end{equation}
where $n$ and $N$ are the number of charge sites on the solute and
solvent molecules, respectively, and $\lambda_i$ is a scaling factor
henceforth called the charge fraction.  Note that this interaction potential
depends on the choice of charges assigned to the solvent and solute
molecules.  In our calculations, the charges on the solute were fitted using the
Kollman-Singh procedure\cite{KollmanSingh} from the gas-phase electron
distribution of the solute.  The use of charges fitted from the
gas-phase calculation of the electron distribution does not account
for the polarization of the solute by the solvent.  To partially compensate for the
neglect of this effect, the total charge on the solvent molecules used
in Eq.~(\ref{electroSoluteSolvent}) is set to the sum of the permanent
and induced charges on the solvent, $q=q_p+q_{in}$.  This is in contrast
to the calculation of the electrostatic interaction between solvent
molecules in which the effective charges are set to $q=q_p+q_{in} /2$.
This approximation is consistent with first-order perturbation theory
in which the polarization energy of the solute by the solvent is
approximated by the polarization energy of the solvent by the
solute\cite{Thesis}.

Note that the only difference between the potentials \( E_{i}({\mathbf{r}}^{s},{\mathbf{r}}^{S}) \)
for different electrostatic indices \( i \) consists of the form of the electrostatic interaction
energy \( E_{i}^{sS}({\mathbf{r}}^{s},{\mathbf{r}}^{S}) \) between the
solute and the explicit solvent atoms. 
The energy \( E^{sS}_{1}({\mathbf{r}}) \) is chosen to be zero in this
work implying that $\lambda_1=0$, 
which amounts to turning off solute charges.  For this system all
electrostatic interactions between the solvent and the solute are
scaled to zero although the solvent and solute still interact through
Lennard-Jones potentials.  By design, the last electrostatic index
is set to unity, $\lambda_{i_{max}}=1$ so that
the energy \( E^{sS}_{i_{max}}({\mathbf{r}}) \) corresponds
to calculating the electrostatic interaction
between the solvent and solute atoms using the fitted configuration-dependent
solute charges obtained \emph{via} the Kollman-Singh method. The additional dimension
of the solute plus solvent state space represented by the
electrostatic index \( i=1,\cdots ,i_{max} \) 
ensures that the solvent configuration adapts smoothly to the solute
via a stepwise process in which the charges of the solute atoms
gradually interact with the other charges in the system.  It should be
noted that an equivalent result can be achieved by using a parallel
tempering scheme in which the label \( i=1,\cdots ,i_{max} \) corresponds
to a stepwise decrease of a ``sampling temperature''
associated with the solute-solvent electrostatic interaction.

In our implementation of the importance sampling, a Markov chain of
extended state space configurations was generated by two types of
transitions.  In transitions of the first type the electrostatic index
\( i \) was kept fixed while the configuration of the system 
\textbf{\( {\mathbf{r}}=({\mathbf{r}}^{s},{\mathbf{r}}^{S}) \)} was
updated using a transition matrix which leaves \( p_{i}({\mathbf{r}})
\) invariant.  The way in which the configurations were updated for
fixed electrostatic index was dependent on the electrostatic scaling
factor.  When the scaling factor was zero both the configuration of
the solute and the solvent were updated simultaneously using the MMBIF
approach.  The background molecular mechanics simulations used to
guide the updates were run so as to produce effectively independent
but energetically reasonable configurations of the entire system.
The simultaneous update of both solute and solvent configuration is
possible in the absence of electrostatic interactions between the
solute and solvent since the fitted charges are
not used in the molecular mechanics auxiliary chain.
However, after the calculation of the \emph{ab initio} DFT energy
in the acceptance-rejection step of the MMBIF method,
the fitted charges for the given solute configuration
can be calculated. When the electrostatic scaling factor is nonzero
and solute-solvent electrostatic interactions occur, the solvent was
allowed to adjust to the charge distribution on the solute by updating
the solvent configuration while maintaining the configuration of the
core reactive region unchanged.  

Transitions of the second type move the system through the auxiliary
parameter space by applying a transition matrix that changes the
electrostatic index while leaving the configuration of the solute and
solvent unchanged.  In this study, the method for
updating the electrostatic index \( i \) consisted
of using a Metropolis algorithm, with a proposal distribution in which
the proposed indices \( i_{\textrm{new}}=i_{\textrm{old}}+1 \) and \( i_{\textrm{new}}=i_{\textrm{old}}-1 \)
are equally probable. The proposed change of electrostatic index is rejected if \( i_{\textrm{new}} \)
is outside the valid range \( i=1,\cdots ,i_{max} \), otherwise it is accepted
with probability 

\begin{equation}
\label{AcceptanceCriterion}
\min \left[ 1,\frac{p\left( i_{\textrm{new}},{\mathbf{r}}\right) }{p\left( i_{\textrm{old}},{\mathbf{r}}\right) 
}\right] =\min \left[ 1,\frac{w_{i_{\textrm{new}}}}{\textrm{w}_{i_{\textrm{old}}}}
\exp \left( -\beta \left( E_{i_{\textrm{new}}}-E_{i_{\textrm{old}}}\right) \right) \right] .
\end{equation}
The marginal distribution of \( i \) with respect to the equilibrium
distribution is given by

\begin{equation}
\label{Choicew}
p(i)=\int p(i,{\mathbf{r}})d{\mathbf{r}} = \int w_{i}p_{i}({\mathbf{r}})d{\mathbf{r}}=w_{i}Z_{i},
\end{equation}
where the configuration partition function \( Z_{i} \) is defined
as

\begin{equation}
\label{Partition}
Z_{i}=\int \exp \left( -\beta E_{i}({\mathbf{r}})\right) d{\mathbf{r}}.
\end{equation}
If the distributions \( p(i,{\mathbf{r}}) \) are all to play a useful
role in sampling, the weights \( w_{i} \) should be chosen such that
a roughly uniform distribution over \( i \) is obtained. Since the
\( Z_{i} \) are initially unknown, suitable values for the weights
are found through a trial and error process using preliminary runs.
To do this, an iterative procedure can be used in which the Markov
chain is simulated using the current values for \( w_{i} \), and
the frequencies \( f_{i} \) with which each distribution is visited
are recorded. 

Next, new and improved weights \( w_{i'} \) are calculated as \( w_{i'}=w_{i}/f_{i} \)
for electrostatic index \( i \). If some of the frequencies
\( f_{i} \) are zero, various elaborations of the estimation procedure
can be used and some of them are summarized in reference {[}\citeonline{Geyer1995}{]}.
The number \( i_{max} \) of values of the electrostatic scaling
parameter $\lambda_i$ used and the actual values of \( w_{i} \),
\( i=1,\cdots ,i_{max} \) are chosen by minimizing the average computer
time necessary for a new solute configuration to appear with an
electrostatic scaling parameter $\lambda_{i_{max}}=1$.  A good
starting estimate for the numbers \( w_{i} \) and
\( i_{max} \) can be obtained by optimizing the weights and the number
of intermediate chains using only the molecular mechanics guiding
potential with a reasonable set of atomic charges on the solute atoms.

\section{Results }

To explore the issue of how importance sampling methods can
be effectively utilized in simulations of reactions in condensed
phases, two different implementations of the MMBIF sampling method
were analyzed.  In the first implementation, the dependence of the solute
charges on the reactive state of the system was taken to vary linearly
with the control parameter $b$ defined in
Eq.~(\ref{a_b_1andb_2Definitions}).  For this simulation, the
electrostatic interactions between the interpolated charges on the
solute and the charges in the solvent were taken to be Coulombic.  In
the second simulation, the simulated tempering method described above
was applied to the solvated acetic acid-methanol cluster.  Both
simulations generate chains of states asymptotically distributed
according to a Boltzmann distribution based on Eq.~(\ref{InteractionSoluteSolventCorrect}) in which
the solute-solvent electrostatic interactions are modeled by
calculating the Coulomb interaction between the gas phase electron
distribution with the charges in the solvent. In both simulations, the desired distribution
for the chain of states based on the universal Kohn-Sham functional
$F[\rho_s (x)]$ can be recovered at the end of the simulation by
re-weighting each of the $N_T$ total configurations by a configuration dependent factor
\begin{equation}
\label{reweight}
W (x_i) = \frac{ e^{-\beta \Delta E_{pol}(x_i)}}{\displaystyle{\sum_{i=1}^{N_T}e^{-\beta \Delta E_{pol}(x_i)}}},
\end{equation}
where
\begin{equation}
\label{DeltaEpol}
\Delta E_{pol} (x_i) = F[\rho_s(x_i)] - \left( F[\rho_0 (x_i)]+V
\right) = F[\rho_s(x_i)]-E^{s}(r_{i}^{s}) - E^{sS}(r_{i}^{s},r_{i}^{S})
\end{equation}
is the difference in the polarization energy of the solute by the
solvent estimated by calculating the energy of the ground state
electron distribution in the presence of the solvent charges and the
energy of a gas-phase electron distribution interacting with the
optimized solvent charges.  

In all simulations, the calculations for
the solvated cluster (shown in Fig. 3)
have been carried out by treating all nuclei as classical point particles. 
The calculations were conducted in the isobaric-isothermal ensemble at \( p=1 \) atm and \( T=298 \) K.
In order to improve the sampling along the reaction coordinate, an
umbrella potential\cite{UmbrellaSampling} was constructed for the
guiding potential using a self-adaptive scheme.  Simulations biased by
the converged umbrella potential yielded a uniform sampling of the
important regions of the reaction coordinate even though the
activation barriers for the proton transfer reaction was on the order
of $28$ $k_{B}T$.

Two thirds of the \emph{ab initio} Markov chain transition steps in
the simulation using the linearly-interpolated charges on the core were generated using the
MMBIF method. The same fraction of base transitions were generated
using the MMBIF updates when the electrostatic index is zero in the
simulated tempering simulation (see
Fig. 4). The rest of the \emph{ab initio} Markov chain transitions
were performed utilizing Metropolis single-variable updates using
the \emph{ab initio} DFT energy. As demonstrated in a previous study
{[}\citeonline{Iftimie1}{]}, the role of the Metropolis DFT updates
is to prevent the guiding molecular mechanics Markov chain of states
from spending a large number of successive steps in those regions
of the configuration space where the molecular mechanics density of
states in the solute configuration space substantially underestimates
the corresponding \emph{ab initio} DFT density of states.

In the previous studies of the formic acid-water\cite{Iftimie1} and
of the malonaldehyde\cite{Iftimie2} systems it was demonstrated that
a useful strategy for optimizing the MMBIF method was to separate
the variables to be updated in a classical MC step into several groups,
with strongly correlated variables grouped together. In the case of
the double proton transfer in the cyclic cluster formed by acetic
acid and methanol, the vibrations of the two methyl groups should
be relatively uncoupled from the motions of the other atoms in the
cluster. Applying this separation of which variables are updated in
the MMBIF method, the percent of rejections of proposed
configurations obtained with the MMBIF method was about \( 30\% \).
The percent of rejections of the transitions which employed single-variable
Metropolis updates using the \emph{ab initio} potential was \( 45\% \).

It is important to compare the computational effort of performing
liquid-phase versus gas-phase simulations of chemical reactions 
within the MMBIF approach.
The potentials of mean force obtained using the reaction coordinate 

\begin{equation}
\label{ReactionCoordinateMalonaldehyde2}
\xi =b=d_{\textrm{O}_{7}\textrm{H}_{4}}-d_{\textrm{O}_{7}\textrm{H}_{6}},
\end{equation}
describing the double proton-transfer in the acetic acid-methanol system
in tetrahydrofuran obtained in the first simulation and in a gas-phase
simulation are represented in Fig. 5. 
Although equal cpu times were spent in computing the two potentials
of mean force depicted in Fig. 5,
the statistical uncertainties for the activation energy calculated for the double
proton reaction rate in the solvated system is approximately \( 2 \)
times larger than the error bar for the activation energy calculated
from the gas-phase simulation. This decrease in the statistical resolution
of the computations in the solvated system with respect to the gas-phase
system is due to a larger integrated correlation time of the overall
Markov chain as well as to
the fact that \( 50\% \) of the cpu time in the simulation of the
solvated system is dedicated to calculating the molecular mechanics
interactions between the \emph{solvent} molecules.  Although a single
\emph{ab initio} calculation is roughly $4$ orders of magnitude slower than a
single update of the configuration of the molecular mechanical
solvent, the long dielectric relaxation of the solvent required that
many updates be carried out on the solvent before an independent
solvent configuration was generated.  This translated into an
approximately equal amount of CPU time for the molecular mechanical
and quantum mechanical components of the simulation.

The approximation of the variation of the charges fitted using the
Kollman-Singh procedure with the solute configuration by a linear
variation with the parameter \( b \) in Equation (\ref{a_b_1andb_2Definitions})
proved quite accurate. The standard deviation of the differences between
the electrostatic interaction energies calculated using these fitted
charges obtained via the Kollman-Singh procedure and the same energies 
calculated using solute charges which
vary linearly with the parameter \( b \) defined in Equation (\ref{a_b_1andb_2Definitions})
is \( 0.6 \) kcal/mol. This small deviation should be contrasted
with the value of \( 4 \) kcal/mol representing the standard deviation
of the values of the electrostatic interaction energy between the
solute and solvent molecules (see Fig. 6).

The distribution of the values of the polarization
energy difference \( \Delta E_{\textrm{pol}} \) defined in Eq.~(\ref{DeltaEpol})
relevant for the final re-weighting of data points is plotted in Fig. 7.
As expected, the values
of \( \Delta E_{\textrm{pol}} \) are distributed over a small range
of energy values of approximately $0.4$ kcal/mol. In fact, calculation
of the activation energy in the double proton reaction rate in tetrahydrofuran
without re-weighting yields an activation energy which differs from
the activation energy in Fig. 5
only by a statistically irrelevant value of $0.1$ kcal/mol.
It should be emphasized that the small values of \( \Delta E_{\textrm{pol}} \)
in Fig. 7 are in part a consequence of approximating the polarization energy
of the solute by the solvent by the polarization energy of the solvent by the solute.
We estimated that if the solvent charges used in Eq. (11) were set to $q^{S}=q^{S}_p+q^{S}_{in}/2$, 
the standard deviation for \( \Delta E_{\textrm{pol}} \) would have been approximately $1.5$ kcal/mol.

The potentials of mean force for the double proton transfer reaction in
tetrahydrofuran calculated using the linearly-interpolated charge approach
and using the simulated tempering method are identical
within error bars. However, if the error bars for the activation energy
in the linearly-interpolated charge method is $0.4$ kcal/mol, they are roughly four times
larger in the simulation using the parallel tempering algorithm for
comparable cpu times. 

The optimal number of electrostatic indices in the simulated tempering
approach was found to be $i_{max}=4$. In the optimized setting
for the simulated tempering method in which equal amounts of time
were spent at all values of the electrostatic index, 
the Gibbs free energy differences $\Delta G_{i,i+1}$
computed for consecutive values of the index were found to be approximately $0.7$
kcal/mol. The Gibbs free energy $G(\lambda)$ of the electrostatic interaction between the 
solute and solvent charges have been estimated using Equation(\ref{Partition})
and are plotted in Fig. 8 as a function of the
fraction of solute charge $\lambda_i$ being turned on. 

Several comments are worth making about the results in Fig. 8.
First, the partition functions $Z_{i}$ for the umbrella potential-biased
ensemble from which the free energies were computed correspond to a 
particular value of the electrostatic fraction parameter
$\lambda_i$.  It should also be noted that although a difference in free energy between
adjacent values of the electrostatic index of the order of $k_{B}T$ seems to suggest a low probability
of rejecting swaps between consecutive values of the index, the actual rejection rate 
was found to be significantly higher.
The relatively large rejection rate is due to the fact that differences in free energy reflect the
\emph{average} energetic \emph{and} entropic differences
between thermodynamic states with different $\lambda $,
whereas the acceptance probability in Equation (\ref{AcceptanceCriterion}) 
involves only the difference
in the energies of the actual configurations to be swapped. In particular, it
was observed that the enthalpic $H(\lambda)$ and entropic $-TS(\lambda)$ contributions 
to the Gibbs free energy $G(\lambda)$
vary in opposite directions as the fraction of the solute charge $\lambda$
increases from zero to one.
These variations of enthalpic and entropic terms with the charge fraction
can be visualized by comparing the results plotted in 
Fig. 8 and Fig. 9.
From the two figures it appears that $H(\lambda)$ increases and $-TS(\lambda)$
decreases with $\lambda$. The opposite directions in which enthalpic and entropic
terms vary with $\lambda$ is reflected in the mobility of the state $(i,{\mathbf{r}})$ of the
simulated tempering algorithm in the $i$ (or $\lambda$) subspace.
Turning our attention to Equation (\ref{AcceptanceCriterion}), note that
an increase of $-TS(\lambda)$ with $\lambda$ suggests that for a large fraction of 
configurations (i,{\textbf{r}}) transitions in which $i$ is \emph{increased} are
accepted only if the ratio
\begin{equation}
\label{r}
w_{\uparrow}(i)=\frac{w_{i+1}}{w_{i}}
\end{equation} 
is approximately unity, $w_{\uparrow}(i) \approx 1$. Such is the case for an
important number of transitions from electrostatic index $i=1$ to
electrostatic index $i=2$, for example, for which an increase in $\lambda$ is accompanied
only by a small decrease and occasionally even an increase in energy.
On the other hand, for a $\lambda$ near unity, a decrease of $H(\lambda)$ with $\lambda$ suggests that
a large fraction of transitions in which $i$ is \emph{decreased} are rejected
unless $w_{\uparrow}(i) \ll 1$. Such is the case in a large number of transitions which are attempted from
the final index $i_{max}$ to $i_{max}-1$ for example, which are accompanied by a significant 
increase in energy.

The choice of the weights $w_{i}$ suggested in Equation (\ref{Choicew}) represents
a good compromise in the sense that transitions which increase and which decrease
$i$ have an equal probability of being accepted on average. Nevertheless, this
analysis points to the fact that if one uses the simulated tempering approach for
the study of chemical reactions in solution, one should
try to avoid turning on the charges of the reactive core all the way
from zero to their final values. As enthalpic and entropic terms will always vary in opposite directions
during the charging process, the sampling could become quite inefficient, especially
when studying reactions in which there is a difference in the net charge between
reactants and transition states, and not just in their dipole moments as in the present
case.

However, this conclusion does not mean that the MMBIF approach used in
the simulation with linearly interpolated charges 
will always be more efficient than the simulated tempering method. The efficiency
of the MMBIF approach relies heavily on the appropriateness of the postulated variation of the
charges in a reactive system with the reaction coordinate. In the present study,
the approximation of the variation of the fitted Kollman-Singh charges
with the solute configuration by a linear 
variation with the parameter \( b \) in Equation (\ref{a_b_1andb_2Definitions})
proved quite accurate. However, such a simple relation could break down, especially
when studying reactions with a net transfer of charge. In this case, we suggest
using an approach which combines the benefits of both methods illustrated
in the present study: the use of a simulated tempering approach in which
the solute charge is gradually modified from a linear variation with the reaction coordinate
to their actual values obtained via the Kollman-Singh approach.

\section{Discussion and Conclusions}

In this article, two important issues on the applicability of the
molecular mechanics based importance sampling method to the study of
reactive events in condensed phase environments have been addressed.
One major concern in developing a successful implementation of the
MMBIF approach is the ease of development of a sufficiently accurate
molecular mechanics potential to guide the sampling.  Although a fully
automated approach to generating guiding potentials for general
reactions is still not available, it is encouraging to note that the use of the same
principles of bond-evolution theory\cite{Krokidis} as in our earlier study of the
malonaldehyde system\cite{Iftimie2} were adequate for designing a molecular mechanics
potential for the acetic acid-methanol system.  This success is
particularly impressive in light of the fact that the malonaldehyde and acetic
acid-methanol systems differ substantially not only in their barrier
heights (by a factor of $4$), but also in the qualitative nature of
the chemical event (one proton versus two protons transferred).
It suggests that bond-evolution theory guidelines are likely to be practical in developing molecular
mechanics potentials for other proton transfer reactions, and for possibly
other types of chemical events.

The study of the double proton transfer reaction between acetic acid
and methanol in tetrahydrofuran also demonstrates that the MMBIF
method can be applied in reaction rate calculations
of chemical transformations in solvents of medium polarity.  In
particular, an increase in the CPU time of factors  of $4$ and $15$
with respect to gas-phase calculations were obtained using two different
sampling methods. Hence, we conclude that it should be possible in many instances
to compute solvent mediated reaction rates with statistical accuracies
comparable to those obtained in gas-phase calculations, even though the complexity
of the calculation is increased enormously by the presence of the solvent.

It should be emphasized that the contribution of the solvent to the
total activation energy in the present study is only on the order of a
few factors of $k_{B}T$ at room temperature.  As a result, the actual
mechanism of the proton transfer event is virtually unchanged from the
process in gas phase.  This simplification allowed the separation of
the reactive \emph{ab initio} core and the solvent degrees of freedom 
into effectively disjoint sets which were updated in isolation in the
parallel tempering implementation of the sampling.
In many cases, the structure of the solvent plays a more substantial
role in the reactive process.  Under such circumstances, care must be
taken to devise methods in which the reactive core and the solvent
structure are updated in a more correlated fashion.  For example, for
a reaction in a solvent with a larger dielectric constant, the
simulated tempering approach can be implemented by incorporating simultaneous
MMBIF updates of the solute and solvent degrees of freedom at each
electrostatic index.

It is informative to compare and contrast the advantages and disadvantages
of using the MMBIF method versus using present day molecular dynamics
based methods for calculating reaction rates in solution using a hybrid
QM/MM approach. The practical use of the simple molecular dynamics
methods for sampling configurations of a system containing a reactive
core which is described using \emph{ab initio} electronic structure
methods and the solvent molecules which are described using a molecular
mechanics potential is limited by the fact that calculating the time-evolution
of the system necessitates the computation of the time-consuming
\emph{ab initio} forces acting on the core atoms. These time-consuming
calculations must be performed in traditional molecular dynamics calculations
even when only the solvent degrees of freedom change significantly. 

Several methods have been proposed for circumventing the inefficiency
of the traditional QM/MM molecular dynamics calculations which are
based on an ``artificial'' separation of time scales
associated with the solute and solvent atoms\cite{TuckermanJCP2002,RothlisbergerJPCB2001}.
The central idea of these methods is to use a large mass in conjunction
with a high temperature thermostat for the solute atoms, whereas the
solvent atoms have usual masses and are in contact with a room temperature
thermostat\cite{TuckermanJCP2002,RothlisbergerJPCB2001}. The large
mass of the core atoms is chosen in such a way that the solvent degrees
of freedom relax on a time scale which is much smaller than the time
scale of the massive core atoms. Multiple-time scale arguments can
be utilized to demonstrate that the integration of Hamilton's equations
describing the evolution of the solute plus solvent system can be performed
by computing the forces acting on the solvent degrees of freedom significantly
more often than the \emph{ab initio} forces acting on the solute without
altering the asymptotic Boltzmann distribution of the configurations
of the system. In addition, the decoupling between the solute and
solvent time scales enables the use of a large temperature thermostat
coupled to the solute atoms without introducing an irreversible heat
flow, thereby ensuring that the average kinetic energy of the core
atoms is comparable with the magnitude of the reaction barrier which
separates reactant and product configurations on the potential energy
surface. Therefore, a relatively small number of \emph{ab initio}
calculations must be performed before a reactive event occurs. 

Both the MMBIF and the modified molecular dynamics approach succinctly
described above have a number of advantages and shortcomings when
studying chemical reactions in solution which are essentially driven
by fluctuations in the structure of the \emph{solute} using a QM/MM
approach. The main disadvantage of the MMBIF method consists of the
fact that a reasonable molecular mechanics description of the reactive
event in the \emph{gas-phase} solute is required, and this molecular
mechanics potential must usually be created from scratch. Nevertheless,
once such molecular mechanics potential has been constructed, the
cpu time necessary for calculating reaction rates in solution is essentially
independent of the characteristics of the solvent such as its dielectric
relaxation time. In contrast, the molecular dynamics approach does
not necessitate prior information with respect to the potential energy
surface of the cluster, although if such information exists, it can
be used to improve the efficiency of the sampling\cite{TuckermanJCP2002,RothlisbergerJPCB2001}.
However, the average time needed to observe a reaction event increases
as the square root of the effective mass of the reactive degree of
freedom. On the other hand, given the requirement
of the separation of time scales between the solvent and solute degrees
of freedom, the characteristic time scale of the solute atom motion
and therefore the lower bound for the value of the mass needed for
the core atoms is \emph{determined} by the duration of the solvent
dielectric relaxation time. Therefore, it appears that the efficiency
of the above mentioned molecular dynamics scheme decreases with the
increase in the solvent dielectric relaxation time. 

The methodology proposed in the present work can be combined with the ideas
presented in reference {[}\citeonline{Iftimie2}{]} to include nuclear quantum effects 
via centroid transition state theory with a supplementary increase in the cpu time
by a factor of $2$ times the number of path-integral beads. Such a
combination of procedures provides a rigorous and practical platform
for calculation of kinetic isotope effects. An important goal for future
work is to clarify the mechanistic origin of the relation between the breakdown of the 
rule of geometric mean in multiple proton transfer reactions and tunneling effects.

\acknowledgements
This work was supported by a grant from the Natural Sciences and Engineering
Research Council of Canada.  R. Iftimie would also like to thank the Ontario
Ministry of Education for financial support.

\newpage
\begin{center}
Figure Captions
\end{center}
Figure 1:  The formation of two cyclic clusters involving one or
two acetic acid molecules and one molecule of methanol. The double-proton 
transfer reaction studied here
takes place along the hydrogen bonds of the cyclic cluster formed from 
one molecule of acetic acid and one molecule of methanol.

\vspace{.1in}
\noindent
Figure 2:  A pictorial view of the QM/MM/continuum solvent
method and of the reaction field approach. The ``reactive
core'' region contains a quantum representation (i.e: nuclei
+ electrons) of the atoms which are involved in the actual covalent
bond-breaking and bond-forming events. The ``explicit solvent''
region contains an atomic representation of the first few shells of
solvent molecules. The effects of the solvent molecules which are
far from the reaction center are included in an implicit manner in
the QM/MM/continuum approach using the reaction field method. In our
implementation, the reaction field method consists of calculating
an effective electrostatic interaction between the dipole moments
of the molecules inside the first two regions (see Equation {[}\ref{ReactionFieldEnergy}{]}).

\vspace{.1in}
\noindent
Figure 3: The structures of the reactant, transition
state and product in the gas-phase double-proton transfer reaction.

\vspace{.1in}
\noindent
Figure 4: A schematic representation of
the simulated tempering method which uses the MMBIF approach to generate
configurations of the system in a simulation where solute polarizability
is neglected.  The values of the electrostatic scaling parameter $\lambda_i$ for electrostatic indices $i=1 \cdots
i_{max}$ are gradually increased from zero to one.

\vspace{.1in}
\noindent
Figure 5:  The calculated potentials of
mean force for the double-proton transfer reaction in the acetic acid-methanol
cluster in gas-phase and in a solution of tetrahydrofuran using the
reaction coordinate \protect\( \xi \protect \) defined in Equation
(\ref{ReactionCoordinateMalonaldehyde2}). Note that the difference
between the activation energies is approximately \protect\( 0.8\protect \)
kcal/mol. This difference is larger than the width of the \protect\( 75\%\protect \)
confidence intervals for the activation energies, which have been
estimated to be \protect\( 0.2\protect \) kcal/mol for the gas-phase
and \protect\( 0.4\protect \) kcal/mol for the liquid-phase
simulations (inset).

\vspace{.1in}
\noindent
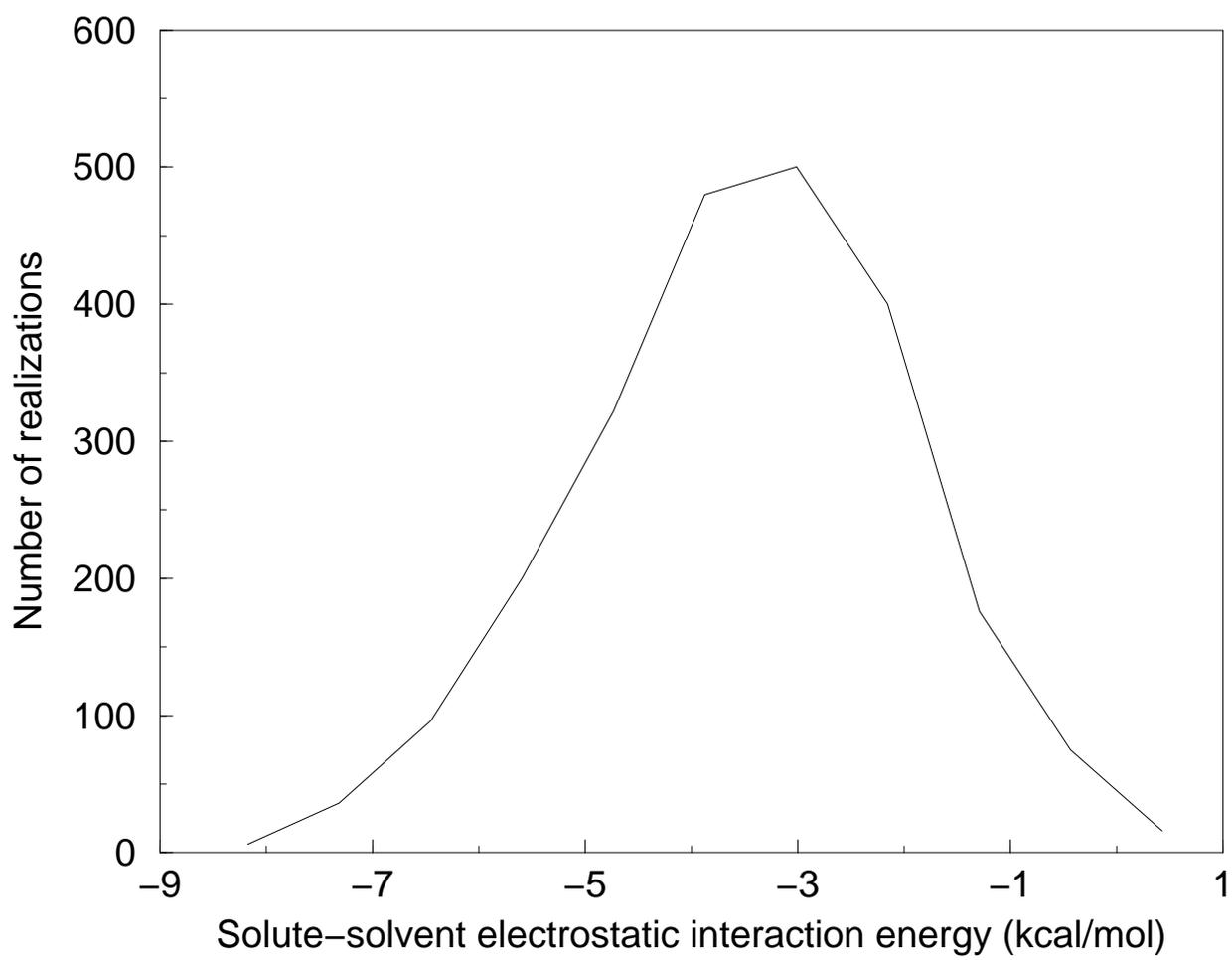
Figure 6:  The values of the
electrostatic interaction energy between the solute and solvent molecules
obtained for $\lambda=1$ in the simulated tempering method in which
the solute charges are the gas-phase Kollman-Singh charges. Note that
the values of the solute-solvent electrostatic interaction energies
are scattered over an energy range of approximately \protect\( 4\protect \)
kcal/mol. In contrast, the differences in the electrostatic interaction
energies calculated using the Kollman-Singh charges and the same energies
calculated using solute charges which vary linearly with the parameter
\protect\( b\protect \) defined in Equation (\ref{a_b_1andb_2Definitions})
are scattered over an interval of only \protect\( 0.6\protect \)
kcal/mol (data not shown).

\vspace{.1in}
\noindent
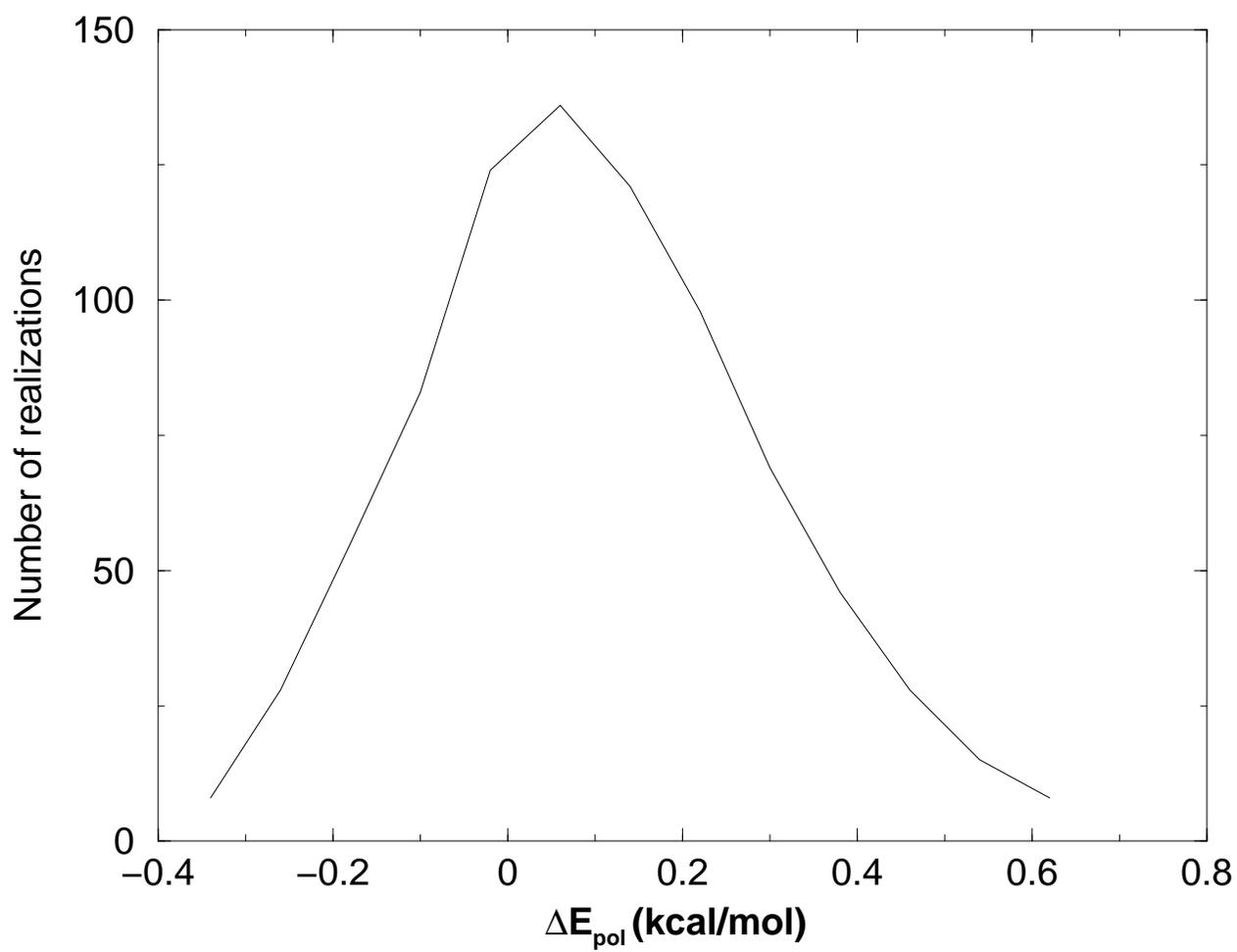
Figure 7:  The values of the
polarization energy difference \protect\( \Delta E_{\textrm{pol}}\protect \)
defined in Equation (\ref{DeltaEpol}). Note
that the distribution of values is centered roughly at \protect\( 0\protect \)
kcal/mol and has a small standard deviation of \protect\( 0.4\protect \)
kcal/mol.

\vspace{.1in}
\noindent
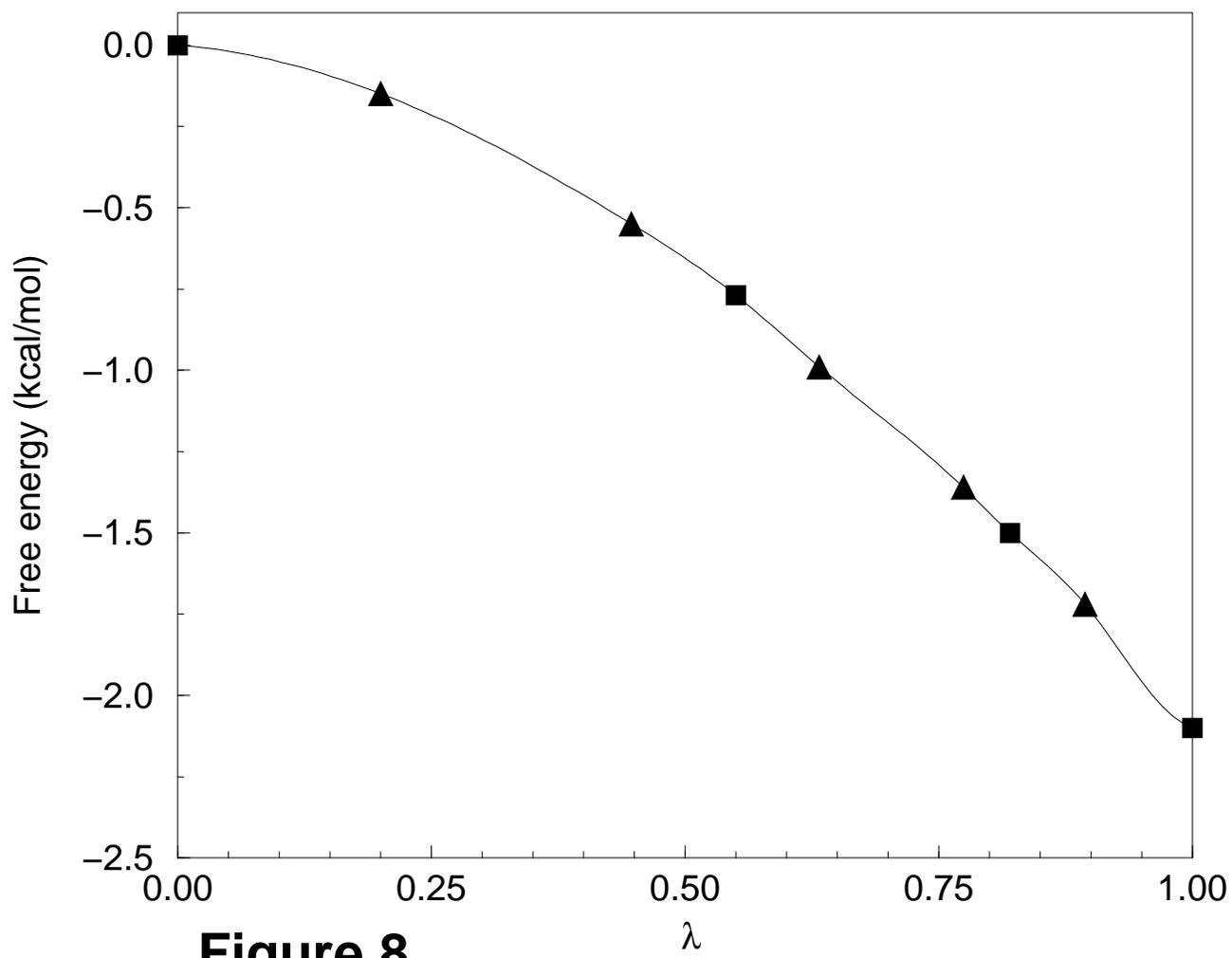
Figure 8:  The free electrostatic interaction energy
between an acetic acid-methanol complex and molecules
of tetrahydrofuran solvent as a function of the charge
state of the complex. $\lambda=0$ and $\lambda=1$ correspond to
fully uncharged and charged solute. The stars and diamonds represent
the estimated values of the free energy obtained from the preliminary
and from the optimized simulated tempering runs, respectively. The continous
line represents a spline interpolation between the computed values.

\vspace{.1in}
\noindent
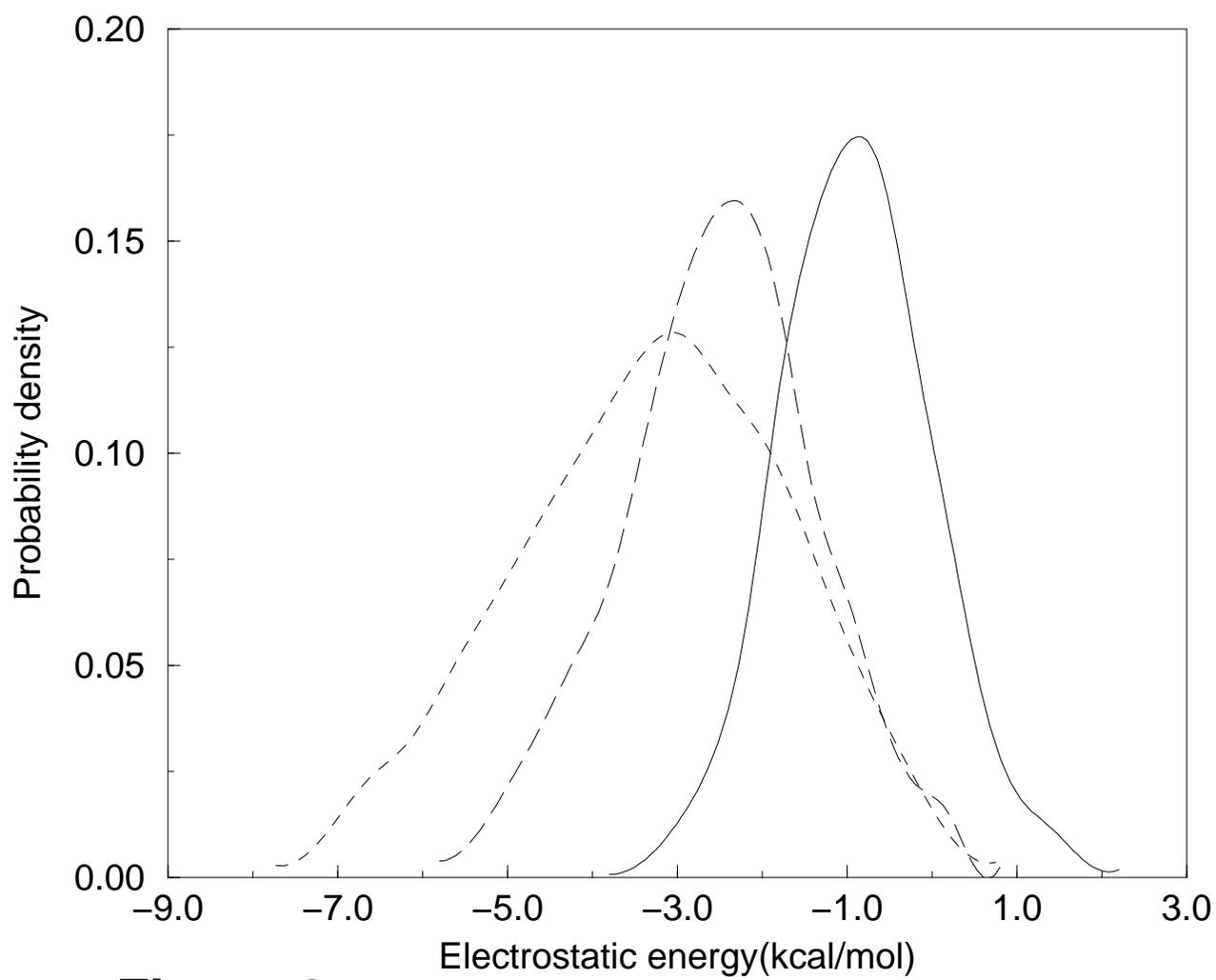
Figure 9:  The normalized probability density
of the solvent-solute electrostatic interaction energy on for
electrostatic indices $i=2$ (solid curve),
$i=3$ (long dashed curve) and $i=4$ (dashed curve). Note that the enthalpies,
calculated as the average solute-solvent electrostatic interaction energies,
are approximately $-1.0$, $-2.4$ and
$-3.5$ kcal/mol. Comparing these results with the corresponding free energies in 
Fig. 8 one obtains the corresponding entropies as
being $0.3$, $1.0$ and $1.4$ kcal/mol. The variations in the mechanical work
$pV$ with the number of chain are negligible.

\newpage
\begin{table}

\caption{\label{OPLSAAExperimentalModOPLSAA}The values of the gas-phase dipole
moment \protect\( \mu _{g}\protect \) expressed in Debyes, and of
the static dielectric constant \protect\( \epsilon _{r}\protect \),
obtained by considering the OPLS-AA charges as permanent charges,
obtained from our modified version of the OPLS-AA force field (see
text) which accounts approximately for electronic polarization effects,
and from experimental data from reference {[}\citeonline{Reichardt}{]}.}

\vspace{0.3cm}

\begin{centering}

\setlength{\extrarowheight}{6pt}

\begin{tabular}{|c|c|c|c|}
\hline 
&
OPLS-AA&
Modified OPLS-AA&
Experimental\\
\hline
\hline 
\( \mu _{g}(D) \)&
\( 1.97 \)&
\( 1.76 \)&
\( 1.75 \)\\
\hline 
\( \epsilon _{r} \)&
\( 6.15\pm 0.3 \)&
\( 7.61\pm 0.38 \)&
\( 7.58 \)\\
\hline
\end{tabular}

\end{centering}
\end{table}

\end{document}